\documentclass[10pt,twocolumn]{article}
\usepackage{cite}
\usepackage{graphicx}
\usepackage{pdfpages}
\usepackage{rotating}
\usepackage{placeins}
\usepackage{csvsimple}
\usepackage{listings}
\usepackage{url}
\usepackage{amsmath}
\usepackage{stmaryrd}

\usepackage{siunitx}
\usepackage{booktabs}
\usepackage{xcolor}
\usepackage{setspace} 
\usepackage{verbatim}

\usepackage{MnSymbol}
\usepackage{wasysym}

\newcommand{\modestsections}{
    \let\section=\subsection
        \renewcommand{\thesubsection}{\arabic{subsection}}
    \let\subsection=\subsubsection  % nos. def. in terms of section as before
    \let\subsubsection=\paragraph
    \let\paragraph=\subparagraph
    \renewcommand{\subparagraph}[1]{\paragraph{##1}
        \typeout{You've used the subparagraph command, which is the same as
                 the paragraph command since you're using modest sections.}}}

\providecommand{\keywords}[1]{\textbf{\textit{Index terms---}} #1}

\begin{document}

%\doublespacing

\modestsections

\title{Creating a Cybersecurity Concept Inventory:\\
A Status Report on the CATS Project}

\author{Alan T. Sherman,$^1$ 
Linda Oliva,$^2$
David DeLatte,$^1$ 
Enis Golaszewski,$^1$
Michael Neary,$^1$\\
Konstantinos Patsourakos,$^1$
Dhananjay Phatak,$^1$
Travis Scheponik,$^1$\\
University of Maryland, Baltimore County (UMBC)\\
Baltimore, Maryland 21250\\
email: \{sherman, oliva, dad, golaszewski, mneary1, konpats1, phatak, tschep1\}@umbc.edu\\
\and
Geoffrey L. Herman,$^3$
Julia Thompson,$^3$\\
University of Illinois at Urbana-Champaign\\
Champaign, Illinois 61820\\
email: \{glherman, jdthomp\}@illinois.edu
}

\date{June 8, 2017}

\maketitle

\footnotetext[1]{Cyber Defense Lab, Department of Computer Science and Electrical Engineering}
\footnotetext[2]{Department of Education}
\footnotetext[3]{Computer Science}
\setcounter{footnote}{3}

\begin{abstract}

We report on the status of our {\it Cybersecurity Assessment Tools (CATS)} project that is creating and validating a
concept inventory for cybersecurity, which assesses the quality of instruction of any first course in cybersecurity.
In fall 2014, we carried out a Delphi process that identified core concepts of cybersecurity.  
In spring 2016, we interviewed twenty-six students to uncover their understandings and misconceptions about these concepts.  
In fall 2016, we generated our first assessment tool—--a draft {\it Cybersecurity Concept Inventory (CCI)}, 
comprising approximately thirty multiple-choice questions.
Each question targets a concept;
incorrect answers are based on observed misconceptions from the interviews. 
This year we are validating the draft CCI using cognitive interviews, expert reviews, and psychometric testing.
In this paper, we highlight our progress to date in developing the CCI.

The CATS project provides infrastructure for a rigorous evidence-based improvement of cybersecurity education.  
The CCI permits comparisons of different instructional methods 
by assessing how well students learned the core concepts
of the field (especially adversarial thinking), 
where instructional methods refer to how material is taught (e.g., lab-based, case-studies, collaborative, competitions, gaming).
Specifically, the CCI is a tool that will enable researchers 
to scientifically quantify and measure the effect 
of their approaches to, and interventions in, cybersecurity education.  

\end{abstract}


\section{Introduction}
\label{intro}

In the coming years, America will need to educate an increasing number of cybersecurity professionals.  
But how will we know if the preparatory courses are effective? Presently there is no rigorous, 
research-based method for measuring the quality of cybersecurity instruction. 
Validated assessment tools are needed so that cybersecurity educators have trusted methods for discerning 
whether efforts to improve student preparation are successful.  
The {\it Cybersecurity Assessment Tools (CATS)} project is developing rigorous instruments 
that can measure student learning and identify best practices.
The first CAT will be a {\it Cybersecurity Concept Inventory (CCI)}
that measures how well students understand basic concepts 
in cybersecurity (especially adversarial thinking) after a first course in the field.

Cybersecurity is a vital area of growing importance for national competitiveness, and 
national reports reveal a growing need for cybersecurity professionals~\cite{Fro2015}. 
As educators wrestle with this demand, there is a corresponding awareness 
that we lack a rigorous research base that informs how to meet that demand. 
Existing certification exams, such as CISSP~\cite{cissp}, are largely informational, not conceptual.  
We are not aware of any scientific studies of any of these tests.  
The CATS Project is developing rigorous assessment tools 
for assessing and evaluating educational practices. 

Since fall 2014, we have been following prescriptions of the National Research Council 
for developing rigorous and valid assessment tools~\cite{Pel01,Pel14,Her2014,Ste2011}.
Our work is inspired in part by the {Force Concept Inventory}~\cite{Hes92}, which
helped transform and improve physics education to employ more active learning methods.

We carried out two surveys using the Delphi process to identify the scope and content of the CATs~\cite{Delphi2016}. 
We then used qualitative interviews to develop a cognitive theory 
that can guide the construction of assessment questions~\cite{FIE2016,interviews2017}.  
Based on these interviews, we developed a preliminary battery of approximately thirty assessment items for the CCI. 
Each item focuses on one of the top five rated concepts (on importance) from our CCI Delphi process. 
The distractors (incorrect answers) for each assessment item are based on student misconceptions observed during the interviews.

In this paper we provide a status report on the CATS project, highlighting our results from the
Delphi process, student interviews, and our development of draft multi-choice questions.
Examples illustrate our method. For more details, see our project 
reports~\cite{Delphi2016,exploring2016,FIE2016,interviews2017}.\footnote{Parts of this paper are drawn from these reports.}

\section{Indentifying Core Concepts through a Delphi Process}
\label{delphi}

In fall 2014, we carried out two Delphi processes aimed at identifying core cybersecurity topics~\cite{Delphi2016}. 
A Delphi process solicits input from a set of subject matter experts to create consensus about contentious decisions~\cite{GOl2010}. 
Topics are refined and prioritized over several rounds, 
where participants share comments without attribution so that the logic of a contributed remark is most significant. 

\begin{table*}
\caption{Top five reconciled CCI topics sorted by median importance (I) and then by
median difficulty (D),
as rated by the Delphi experts using a 1-10 Likert scale where 10 is the greatest.}
\label{tbl:delphi}
\medskip

\centering

\begin{minipage}{0.48\textwidth}
\centering

\begin{tabular}{l|c|c}
{Topic} & I & D \\
\hline %\hline
Identify vulnerabilities and failures & 9 & 8\\
Identify attacks against CIA\footnote{CIA refers to confidentiality, integrity, availability.}
triad and authentication & 9 & 8\\
Devise a defense & 9 & 7\\
Identify the security goals & 9 & 6\\
Identify potential targets and attackers & 9 & 5\\
\end{tabular}
\end{minipage}
\end{table*}

The first process was for the CCI, which is for students completing any first course in cybersecurity.
The second process was for a {\it Cybersecurity Curriculum Assessment (CCA)}, which is for 
students graduating from college about to enter the workforce as cybersecurity professionals.
Here, we focus on the CCI.
We conducted each process electronically, through emails between Delphi leaders and the panel of experts, 
and through web forms to collect survey data.\footnote{The Delphi leaders were Sherman and Parekh, who consulted with Herman
who has notable experience with this process.}
We carried out five rounds for CCI and four for CCA.
To the authors' knowledge, these are the first Delphi processes for cybersecurity to identify core concepts. 

A total of thirty-six experts participated in the initial topic generation phase, including thirty-three for the CCI process. 
The selected experts constitute a diverse group of men and women from over a dozen US states and Canada, 
working as cybersecurity authors, educators, and professionals from industry and government~\cite{Delphi2016}.
Each expert holds a PhD in a cybersecurity-related field and teaches cybersecurity, 
or works as a cybersecurity professional.
The project website lists the experts and their affiliations.\footnote{http://cisa.umbc.edu/cats/}

Experts rated CCI and CCA topics according to three distinct metrics: 
(1)~Importance, (2)~Difficulty, and (3)~Timelessness using a 1--10 Likert scale,
where 10 is the greatest. 
If an expert chose to rate a topic outside the interquartile range, 
they were asked to provide a written justification for their deviation from the consensus. 
These comments enabled dissenting experts to sway the majority.
Once the deadline for the round passed, the Delphi leaders compiled summary statistics for each topic.
These descriptive statistics and data visualization provided the Delphi leaders with information about the level of consensus.

Responses from the first rounds of CCI and CCA were unexpectedly similar; though, 
adversarial thinking was a prevalent theme among CCI responses. 
To ensure that CCI was headed in a distinct direction from CCA, a second topic identification round was performed for CCI only. 
Delphi leaders asked participants to provide topics focused on adversarial thinking, 
which the Delphi leaders and the experts felt constitutes a vital core of cybersecurity. 
The restarted CCI process produced thirty topics.

Table~\ref{tbl:delphi} lists the top five topics from the CCI process sorted by median importance.
The main contribution of this phase of the project is a numerical rating of the importance and difficulty of concepts in cybersecurity.
It is prudent to identify topics that are difficult, since those topics may provide the greatest barriers to mastery. 
These ratings can be used to identify core concepts---cross-cutting ideas that connect knowledge in the discipline---which 
can guide the design of curriculum, assessment tools, and other educational materials and policies.

The results of the Delphi processes, especially the CCA process, identified a range of specialized topics, 
reflecting the broad, multi-faceted aspects of cybersecurity. This range of facets can make prioritizing content 
in cybersecurity education difficult and make it difficult for students to discern how topics connect. 
The five topics rated most important by the Delphi experts in CCI stand out as important and timeless 
concepts that can create priorities in instruction and help students organize their learning.

In addition, these results help clarify, distill, and articulate what is cybersecurity, which this project sees 
as the management of information and trust in an adversarial cyber world.

\section{Uncovering Misconceptions by Interviewing Students}
\label{interviews}

To study how students reason about core cybersecurity concepts, 
in spring 2016, we conducted twenty-six think-aloud interviews with cybersecurity students
who had completed at least one course in cybersecurity~\cite{FIE2016,interviews2017}. 
We recruited these students from three diverse institutions: 
University of Maryland, Baltimore County, Prince George's Community College, and Bowie State University. 
During the interviews, students grappled with security scenarios designed to probe student understanding of cybersecurity, 
especially adversarial thinking. 
No prior research has documented student misconceptions about cybersecurity concepts 
nor how they use adversarial models to guide their reasoning. 

Drawing upon the five topics from the CCI Delphi process, we
developed a set of twelve engaging cybersecurity scenarios, organized into three semi-structured interview protocols.  
These scenarios were not formally vetted by external experts.
Each interview lasted about one hour during which an interviewer asked the student to identify and scrutinize
security concerns for each of four scenarios.  
We audio- and video-recorded each interview and produced a written transcript.  
In addition, we saved any diagrams drawn by the student.
We analyzed student statements using a structured qualitative method, novice-led paired thematic analysis, 
to document student misconceptions and problematic reasonings~\cite{Her2012,Mon2015}. 

The following excerpt illustrates a typical exchange that took place during an interview.
In the given scenario, the interviewer presents evidence of a possible SQL injection
attack vulnerability, one of the most common software security issues.

\bigskip
\noindent {\bf Scenario A3: Database Input Error.}\\
{\it Interviewer 1:} When a user Mike O'Brien registered a new account for an online shopping site, 
he was required to provide his username, address, first and last name.  
Immediately after Mike submitted his request, you—--as the security engineer—--receive a database input error message in the logs.  
What might you infer from this error message?

\noindent {\it Subject:} It's very costly to implement security at the server level.  
But at the client side, SQL injection methods are used by hackers to get into database. 
Simple queries, like avoiding special characters, which can be used for hacking, can be avoided.

\noindent {\it Interviewer 1:} Would it be sufficient for them to add that sort of input checking at the computer?

\noindent {\it Subject:} Client level, yeah. Implementing security at the client level is easier than in database and server level. 
Because at server level the cost is insane.
 
\noindent {\it Subject:} $\dots$ I would say somebody is trying to do SQL? 
When they are trying to gain access to someone's account by adding in a piece of SQL that would always return true, 
but there are ways to do that too.

\noindent {\it Interviewer 2:} How would you do it?

\noindent {\it Subject:} You would not allow certain characters to say username, 
password field or you would make sure that there is enough quotations around the password, 
so the if someone were to try to break out of the quotations whatever content the user entered would be escaped. 
So you wouldn't accent a quote as a raw quote but an escaped character, 
so that the query database system would know that the quote 
is not the end of a set of strings as part of the query its data in the query.
$\blacksquare$ 

\medskip
 
Among other issues, this exchange illustrates the following theme that we commonly observed:
incorrect assumptions---failure to see vulnerabilities and limiting the adversary.  
The student incorrectly asserts that it would be sufficient to implement a defense at the
client side, apparently failing to appreciate that an adversary might attack the server
and/or that the client might be malicious. 

In addition, the student's focus on the client (with less attention to the server) 
suggests a user bias, perhaps because the student's experience may have primarily been as a user.
Such a limited mental perspective can blind the student to vulnerabilities in other aspects of the system
and lead to inappropriate conclusions.

In addition to the themes of incorrect assumptions and biases, we also commonly observed
two additional themes: over generalizations and conflating concepts.  For example, 
some students stated that no Internet communications are secure;  others seemed to believe
that use of biometrics always greatly enhances the strength of authentication systems.
We observed students conflating many concepts, including, for example, 
authorization and authentication, hashing and encryption, and threat and risk.

Biases we observed include user, physical, and personal.  For example, some students
assumed that physical security is usually stronger than software security.  
Others assumed that data security is usually stronger in the USA than in other countries.
These biased assumptions are not necessarily true, and there is not always a clean boundary
separating the referenced categories (e.g., the security of software systems depends in part on the security
of hardware components, and hackers often cross international boundaries electronically).

In a companion paper~\cite{exploring2016}, we discuss answers to six of our scenarios, to
provide instructive case studies for students and educators.

\section{Drafting Assessment Questions}
\label{questions}

Building on the CCI Dephi process, scenarios, and student interviews, we
developed thirty draft CCI multiple-choice questions of varying difficulty.  
For each of the twelve scenarios, we developed one or more questions.  
Each question targets one of the top five topics identified in the CCI Delphi process.
Each question has exactly one best answer, and 
the distractor (incorrect) choices are inspired, whenever possible, by commonly observed
misconceptions from the interviews.

We developed these assessment items through a collaborative iterative process while seated around
a conference table.  Our team included experts in cybersecurity and education, including
professors and graduate students.  
We endeavored to follow best practices for creating multiple choice questions, including 
heeding the ``Vanderbilt advice''~\cite{Vanderbilt}.
It was helpful to revisit draft questions periodically to revise and polish them.
Everyone discovered that producing high-quality questions is challenging.

The representative draft question in Figure~\ref{fig:question}
illustrates the results of our process.
This question targets the topic of devising a defense in the context of the
database input error scenario introduced in Section~\ref{interviews}.

\begin{figure}[h]
\noindent {\bf Scenario A3.}  {\it When a user Mike O'Brien registered a new account for an online shopping site, 
he was required to provide his username, address, first and last name.  
Immediately after Mike submitted his request, 
you—--as the security engineer—--receive a database input error message in the logs.}

\medskip

\noindent {\bf Question A3-3.}  Choose the best defense
to protect against possible security problems suggested by this error:
\begin{description}
\item[A.] Sanitize input at the server side.
\item[B.] Place security controls at the client side.
\item[C.] Require all characters to be from a restricted set of characters.
\item[D.] Implement the system in a secure programming language.
\item[E.] Test the software more thoroughly before deploying it.
\end{description}
\vspace*{-10pt}

\caption{A sample CCI assessment item associated with Scenario~A-3 (database input error).}
\label{fig:question}
\end{figure}

Distractor B is inspired by the misconception discussed for the interview excerpt from Section~\ref{interviews}.
The other distractors are not bad actions to include, but they alone do not mitigate the vulnerability.
Alternative A is clearly the best choice among incomplete alternatives.

\section{Next Steps}
\label{next}

This year we will validate our draft CCI multiple-choice questions 
using cognitive interviews, expert reviews, and psychometric testing. 
First, we will carry out a pilot study seeking feedback from at least twenty
experts and administering the CCI to at least 200 students.  
We will use the results to revise and improve the draft questions.  
Then, applying item-response theory~\cite{Ham91,Ham93}, we will analyze
results from administering the CCI to at least 1000 students.  
We expect the final test to comprise twenty-five questions, five
per topic.

In addition, we plan to continue development of the CCA.  The next step is
to develop draft questions.  We envision that the CCA will target the same
topics from the CCI but at greater technical depth.

The CATS project is meeting the need for rigorous evidence-based assessment tools
to inform the development of best practices in cybersecurity education.
In particular, the CCI will enable researchers to scientifically quantify  
and measure the effect of their approaches to first courses in cybersecurity.
We welcome feedback on our work, and we would be delighted to 
hear from any researcher who might like to participate in our study.

\section{Acknowledgments}
\label{acks}

This work was supported in part by the U.S. Department of Defense under CAE-R grants H98230-15-1-0294 and H98230-15-1-0273, 
and by the National Science Foundation under SFS grant 1241576.

\small
\bibliography{CATS-bib}
\bibliographystyle{alpha}

\bigskip \noindent
Appears in the proceedings of the 2017 National Cyber Summit (June 6--8, 2017, Huntsville, AL).

\end{document}